\documentclass{aa}  
\usepackage{graphicx}
\usepackage{txfonts}
\begin{document}
   \title{Non-radial oscillations in the red giant HR~7349 measured by CoRoT\thanks{The CoRoT space mission has been developed and is operated by CNES, with
the contribution of Austria, Belgium, Brazil, ESA, Germany and Spain.}}

    \author{F. Carrier\inst{1}
           \and
	    J. De Ridder\inst{1}
          \and
          F. Baudin\inst{2}
          \and
          C. Barban\inst{3}
          \and
          A.~P. Hatzes\inst{4}
          \and
          S. Hekker\inst{5,6,1}
          \and
          T. Kallinger\inst{7,8}
	  \and
	  A. Miglio\inst{9}
	  \and
	  J. Montalb\'an\inst{9}
	  \and
	  T. Morel\inst{9}
          \and
          W.~W. Weiss\inst{7}
           \and
	  M. Auvergne\inst{3}
	  \and
	  A. Baglin\inst{3}
	   \and
	  C. Catala\inst{3}
	  \and
	  E. Michel\inst{3}
	  \and
	   R. Samadi\inst{3}
	            }

  \institute{Instituut voor Sterrenkunde, Katholieke Universiteit Leuven, Celestijnenlaan 200D, 
	     B-3001 Leuven, Belgium \\ \email{fabien@ster.kuleuven.be}
              \and 
             Institut d'Astrophysique Spatiale, Campus d'Orsay, F-91405 Orsay, France
	     \and
	     LESIA, UMR8109, Universit\'e Pierre et Marie Curie, Universit\'e Denis Diderot, 
	     Observatoire de Paris, 92195 Meudon, France            
	       \and
              Th\"uringer Landessternwarte, D-07778 Tautenburg, Germany
              \and
	      University of Birmingham, School of Physics and Astronomy, Edgbaston, Birmingham B15 2TT, UK
              \and
              Royal Observatory of Belgium, Ringlaan 3, 1180 Brussels, Belgium
              \and
              Institute for Astronomy, University of Vienna, T\"urkenschanzstrasse 17, A-1180 Vienna, Austria
              \and
	      Department of Physics and Astronomy, University of British Columbia, 6224 Agricultural Road, Vancouver, BC V6T 1Z1, Canada
              \and
	      Institut d'Astrophysique et de G\'eophysique de l'Universit\'e de Li\` ege, All\'ee du 6 Ao\^ut, 17 B-4000 Li\` ege, Belgium
	      }

   \date{Received ; accepted }

 
  \abstract
   {Convection in red giant stars excites resonant acoustic waves whose frequencies depend on the sound speed
inside the star, which in turn depends on the properties of the stellar interior. Therefore, asteroseismology is 
the most robust available method for probing the
internal structure of red giant stars.}
  {Solar-like oscillations in the red giant HR~7349 are investigated.}
   {Our study is based on a time series of 380\,760 photometric measurements spread over 5 months obtained with the CoRoT satellite.
   Mode parameters were estimated using maximum likelihood estimation of the power spectrum.}
   {The power spectrum of the
high-precision time series clearly exhibits several identifiable peaks between 19 and
40\,$\mu$Hz showing regularity with a mean large and small spacing of $\Delta\nu$\,=\,3.47\,$\pm$\,0.12\,$\mu$Hz and 
$\delta\nu_{02}$\,=\,0.65\,$\pm$\,0.10\,$\mu$Hz. 
Nineteen individual modes are
identified with amplitudes in the range from 35 to 115\,ppm. The mode damping time is estimated to be 14.7$^{+4.7}_{-2.9}$\,days.}
   {}


   \maketitle
%

\section{Introduction}

The analysis of the oscillation spectrum provides an unrivaled method for probing the stellar
internal structure because the frequencies of these oscillations depend on the sound speed inside the
star, which in turn depends on the density, temperature, gas motion, and other properties of the stellar
interior. High-precision spectrographs have acquired data yielding to a rapidly growing list of solar-like oscillation detections
in main-sequence and giant stars (see e.g., Bedding \& Kjeldsen \cite{bk07}, Carrier et al. \cite{cel08}). 
In a few years, we have moved from ambiguous detections to firm measurements. Among these, only a few are
related to red giants, e.g., \object{$\xi$~Hya}, Frandsen et al. 
(\cite{frandsen}), \object{$\epsilon$~Oph}, De Ridder et al. (\cite{joris1}), and 
\object{$\eta$~Ser}, Barban et al. (\cite{barban}). 
The reason is that longer and almost uninterrupted time series are needed to characterize the oscillations in
red giants, because of longer oscillation periods than main-sequence stars, and long observing runs are
difficult to obtain using high-accuracy spectrographs.

The CoRoT (COnvection ROtation and planetary Transits) satellite (Baglin \cite{bag}) is perfect for this
purpose because it can provide these data for a large number of stars simultaneously.
The CoRoT satellite continuously collects white-light high-precision photometric observations for 10
bright stars in the so-called {\it seismofield}, as well as
3-color photometry for thousands of relatively faint stars in the so-called {\it exofield}. The primary
motivation for acquiring this second set of data is to detect planetary transits,
but the data
are also well suited to asteroseismic investigations. De Ridder et al. (\cite{joris2})
unambiguously detected long-lifetime non-radial oscillations in red giant stars in the exofield data of CoRoT, which is
an important breakthrough for asteroseismology. Indeed, observations from either the ground or other
satellites have been unable to confirm the existence of 
non-radial modes and determine a clear value of the mode lifetime. Hekker et al. (\cite{hekker}) presented a more detailed classification of the red giants observed by CoRoT.

\object{HR 7349} (HD 181907) is a bright equatorial G8 giant star ($V$ = 5.82) that is an excellent target for asteroseismology. This star was
selected as a secondary target during the first long run of the CoRoT mission.
In this paper, we thus report on photometric CoRoT observations of \object{HR~7349} resulting in the detection and identification of p-mode oscillations.
The non-asteroseismic observations are presented in Sect.~2, the CoRoT data and frequency analysis in
Sects.~3 and 4, and the conclusions are given in
Sect.~5.


\section{Fundamental parameters}

\subsection{Effective temperature and chemical composition}

We used the line analysis code MOOG, Kurucz models, and a high-resolution FEROS spectrum obtained in June 2007 to carry out an LTE abundance study of HR~7349.
The effective temperature and surface gravity were estimated from the excitation and ionization equilibrium of a set of iron lines taken from the 
line list of Hekker \& Mel\'endez (\cite{hekker1}). We obtain $T_{\rm eff}$=4790$\pm$80 K and [Fe/H]=--0.08$\pm$0.10 dex, while the abundance pattern of 
the other elements with respect to Fe is solar within the errors. 
The full results of the abundance analysis will be reported elsewhere (Morel et al., in preparation).
We also determined a photometric temperature given by the relation in Alonso et al. (\cite{alonso}) using
the dereddened color index ($B$-$V$) (see Sect.~\ref{lum})  and found 4704$\pm$110\,K, which agrees with the spectroscopic value.
We finally adopt a weighted-mean temperature of 4760$\pm$65\,K.

\subsection{Luminosity}
\label{lum}
Even for such a bright star, the interstellar extinction in the direction of the Galactic center is not negligible.
From the \textsc{Hipparcos} parallax $\Pi=9.64 \pm 0.34$\,mas (van Leeuwen \cite{vanleeuwen}) and the value of ($B$-$V$)\,=\,1.093 in the \textsc{Hipparcos} catalog, 
an absorption of A$_V$\,=\,0.185\,mag is derived for the region of the star (Arenou et al. \cite{arenou}), which corresponds
to an E$_{B-V}$\,=\,0.052.
Combining the magnitude $V = 5.809 \pm 0.004$ (Geneva photometry, Burki et al. \cite{bu08}),
the \textsc{Hipparcos} parallax, 
the solar absolute bolometric magnitude $M_{\mathrm{bol},\,\odot}=4.746$
(Lejeune et al. \cite{le98}), and the mean
bolometric corrections $BC = -0.40 \pm 0.04$\,mag
($BC = -0.42 \pm 0.06$\,mag according to the calibration of Flower, \cite{flower} and $BC = -0.38 \pm 0.05$\,mag by Alonso et al., \cite{alonso}), 
we find a luminosity for HR~7349 of $L=69 \pm 6$\,$L_{\odot}$.

\subsection{Rotational velocity}
We determined the rotational velocity of the star by means of a spectrum taken with the spectrograph \textsc{Coralie} installed on the 1.2-m Swiss telescope, ESO La Silla, Chile.
According to the calibration of Santos et al. (\cite{santos}), we determined a v\,$\sin i$\,=\,1.0\,$\pm$\,1.0\,km\,s$^{-1}$. 
For this small value of the projected rotational velocity, 
we do not expect to see any split modes in the power spectrum.

\subsection{Large spacing estimation}
\label{lse}
An estimation of the mass of HR~7349 may be obtained by matching evolutionary tracks to the $L$-$T_{\rm
eff}$ error box in the HR diagram. However, in the red giant part of
the HR diagram, this determination is not robust at all. Assuming a mass of between 0.8 and 3\,M$_{\odot}$ and scaling from the solar case (Kjeldsen \& Bedding \cite{kjebed}), 
a large frequency spacing of 2.8-5.5\,$\mu$Hz is expected.

\section{CoRoT Observations}
\label{co}
   \begin{figure}
   \resizebox{\hsize}{!}{\includegraphics{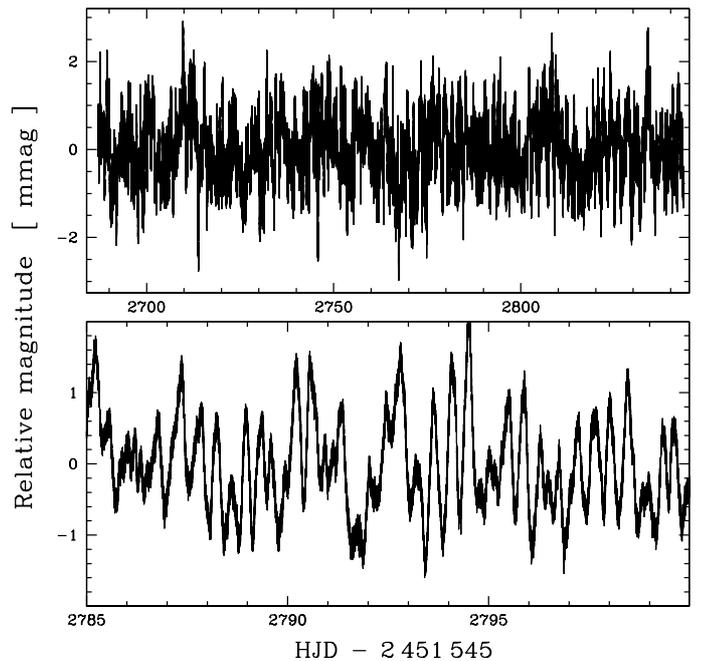}}
   \caption{The total CoRoT light curve (top) and a zoom (bottom) of HR~7349. This light curve is
   detrended with a polynomial fit (order 8), which only affects frequencies below 6\,$\mu$Hz.
   A periodicity of about 8.5 hours can be seen in the zoom, corresponding to oscillation modes close to 30\,$\mu$Hz.}
              \label{light}%
    \end{figure}
\object{HR 7349} was observed with the CoRoT satellite for 5 consecutive months. CoRoT was launched on 2006 December 27
from Ba\"{i}konur cosmodrome on a Soyuz Fregat II-1b launcher. The raw photometric data acquired with
CoRoT were reduced by
the CoRoT team. A detailed description of how photometric data are extracted for the seismology field was presented in Baglin (\cite{bag}).
A summary can be found in Appourchaux et al. (\cite{appourchaux}).

\begin{figure}
   \resizebox{\hsize}{!}{\includegraphics{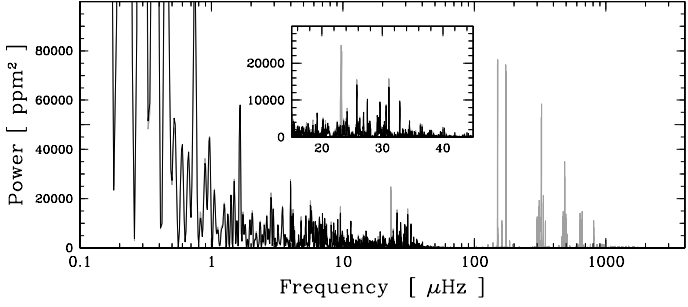}}
   \caption{Power spectra of the original data (grey) and interpolated data (black). The range of the oscillation is zoomed in the inset.
   The interpolation drastically reduces the amplitude of aliases, in particular the one at 23\,$\mu$Hz.}
              \label{figboth}%
    \end{figure}

In the seismofield, CoRoT obtains one measurement every 32 seconds. The observations lasted for 156.64 days from May 11th to October 15th 2007.
The light curve shows near-continuous coverage over the 5 months, with only a small number of
gaps due mainly to the passage of CoRoT across the South Atlantic Anomaly. These short gaps were filled by suitable interpolation (Baglin \cite{bag}), 
without any influence on the mode extraction because it 
only affects the amplitude of frequencies far above the oscillation range of our target (see Fig.~\ref{figboth}). 
The duty cycle for HR~7349 before interpolation was 90~\%. For the frequency analysis (see
Sect.~\ref{fa}), the light curve was detrended with a polynomial fit to remove the effect of the aging of the CCDs (see Auvergne et al. \cite{auvergne}).
This detrending has no consequence
on the amplitude or frequency of oscillation modes, since it only affects the power spectrum for frequencies lower than 6\,$\mu$Hz.
The light curve shows variations of a timescale of 8-9 hours and peak-to-peak amplitudes of 1-3\,mmag (see Fig.~\ref{light}). This signal is a superposition of tens of smaller
modes with similar periods (see Sect.~\ref{fa}).

\section{Frequency analysis}
\label{fa}
\subsection{Noise determination}
\label{nb}
   \begin{figure}
   \resizebox{\hsize}{!}{\includegraphics{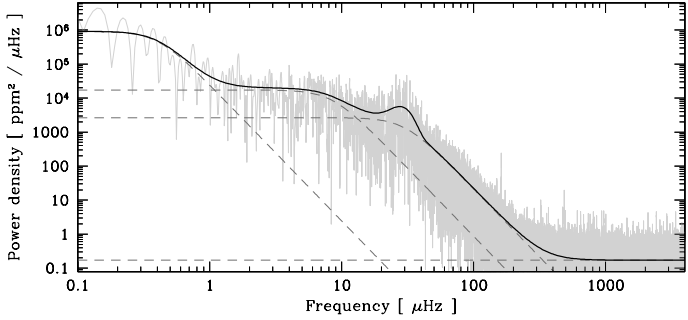}}
   \caption{Power density spectrum of the photometric time series of HR~7349 and a multi-component function (black line) fitted to the
   heavily smoothed power density spectrum. The function is the superposition of three power-law components (dashed lines), white noise (horizontal dashed line)
   and a power excess hump approximated by a Gaussian function.}
              \label{logfit}%
    \end{figure}
\begin{figure}
   \resizebox{\hsize}{!}{\includegraphics{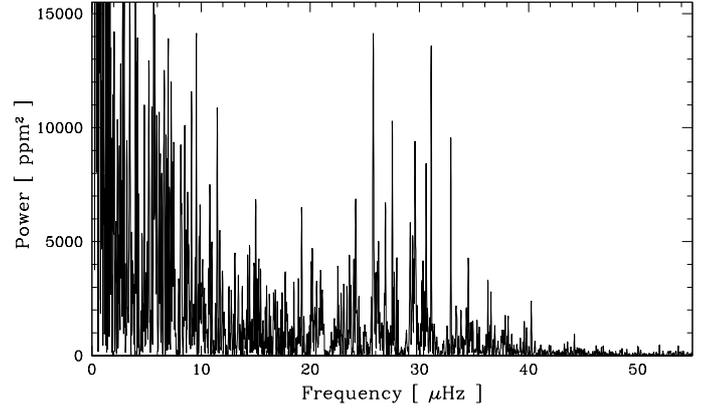}}
   \caption{Power spectrum of the CoRoT observations of HR~7349. Only a polynomial fit was removed from the original data.}
              \label{power}%
    \end{figure}

We computed the power spectra of CoRoT light curves with both gaps and interpolated points (see Sect.~\ref{co}). The resulting
power spectra are quasi-identical, the interpolation not affecting the oscillations but suppressing the
aliases (the most important of which lies at 23\,$\mu$Hz).
We thus analyze the interpolated time series with negligible alias amplitudes.
The time base of the observations gives a formal resolution of 0.07\,$\mu$Hz.
The power (density) spectrum of the time series, shown in Figs.~\ref{logfit} and~\ref{power}, exhibits a series of peaks between 20 and 40\,$\mu$Hz, exactly where the solar-like oscillations
are expected for this star. The power density spectrum is independent of the observing window: this is
achieved by multiplying the power by the effective length of the 
observing run (we have to divide by the resolution for equidistant data), which is calculated to be the
reciprocal of the area beneath the spectral window in power (Kjeldsen et al. \cite{kjel}).
We note that to obtain the same normalisation as in Baudin et al. (\cite{baudin}), we multiply the power by the effective length of the observation divided by four.
Typically for such a power spectrum, the noise has two components:
\begin{itemize}
\item At high frequencies it is flat, indicative
of the Poisson statistics of photon noise. 
\item Towards the lowest frequencies, the power scales inversely with frequency, as expected for instrumental instabilities and noise of stellar origin like granulation. 
\end{itemize}
For the Sun, it is common practice to model the background signal with power laws to allow accurate measurements of solar
oscillation frequencies and amplitudes (Harvey \cite{harvey}, Andersen et al. \cite{andersen}, Aigrain et
al. \cite{aigrain}). To study this "noise", we compute the power density spectrum shown in Fig.~\ref{logfit} and 
fit a smoothed version of this spectrum with a sum of N power laws 
\begin{equation}
P (\nu) = \sum_{i=1}^N P_i =  \sum_{i=1}^N \frac{A_i}{1+(B_i \ \nu)^{C_i}} ,
\end{equation}
where this number N depends on the frequency coverage, $\nu$ is the frequency, A$_i$ is the amplitude of
the $i$-th component, $B_i$ is its characteristic timescale, and C$_i$ is the slope of
the power law. For a given component, the power remains approximately constant on timescales longer than $B_i$, and drops off
for shorter timescales. Each power law corresponds to a separate class of physical phenomena, occurring on a different characteristic
timescale, and corresponding to different physical structures on the surface of the star. In our case, we
fixed the slope to 4,
which is a typical value for the Sun (Aigrain et al. \cite{aigrain}, Michel et al. \cite{michel}). Moreover, this value allows us to fit our power density spectrum well. 

To model the power density spectrum, we added a white noise P$_n$ and a power excess hump produced by the
oscillations, which was approximated by a Gaussian function
 \begin{equation}
P (\nu) =  \sum_{i=1}^N \frac{A_i}{1+(B_i \ \nu)^{C_i}} + P_n + P_g \ e^{-(\nu_{max}-\nu)^2/(2 \sigma^2)} \ .
\end{equation}
The number of components N is determined iteratively: we first made a single component fit, and additional
components were then added until they no longer improved the fit.
In our case, we limited the number of components to three.
We note that for this star, part of the low frequency variation is caused by the aging of the CCD. The timescales for the different noise sources ($B_i$)
are 3.6 $\times$ 10$^6$, 6.9 $\times$ 10$^4$, and 1.1 $\times$ 10$^4$ s. The noise at high frequencies
P$_n$ is only 0.17 ppm$^2 / \mu $Hz and the oscillations are centered on 28.2\,$\mu$Hz.

\subsection{Search for a comb-like pattern}
In solar-like stars, p-mode oscillations are expected to produce a
characteristic comb-like structure in the power spectrum with mode
frequencies
$\nu_{n,\ell}$ reasonably well approximated by the asymptotic
relation (Tassoul \cite{tassoul80}):
\begin{eqnarray}
\label{eq1}
\nu_{n,\ell} & \approx &
\Delta\nu(n+\frac{\ell}{2}+\epsilon)-\ell(\ell+1) D_{0}\;.
\end{eqnarray}
Here $D_0$ (which equals $\frac{1}{6} \delta\nu_{02}$ if the 
asymptotic relation holds exactly) and $\epsilon$ are sensitive to the sound speed near the core and in
the surface layers, respectively. 
The quantum numbers $n$ and $\ell$ correspond to the radial
order and the angular degree of the modes, and $\Delta\nu$ and
$\delta\nu_{02}$
are the large and small spacings.
We note that a giant star such as \object{HR~7349} is expected to show
substantial deviations from its regular comb-like structure described
above (Christensen-Dalsgaard \cite{chris}). This is because some mode frequencies, except for 
$\ell=0$, may be shifted from their usual regular spacing by avoided crossings
with gravity modes in the stellar core (also called `mode bumping') (see e.g., Christensen-Dalsgaard et al. \cite{cd95} and Fernandes \& Monteiro \cite{fm03}).  We
must keep the possibility of these mixed modes in mind when
attempting to identify oscillation modes in the power spectrum. Moreover, the ratio of lifetime to oscillation periods is usually far smaller for red giants 
(because of their longer periods)
than for solar-like stars, which can complicate the mode detection (see e.g., Stello et al. \cite{stello}, Tarrant et al. \cite{tarrant}).

\begin{figure}
   \resizebox{\hsize}{!}{\includegraphics{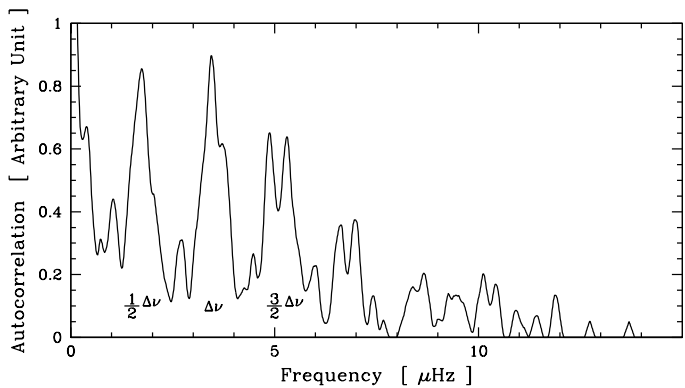}}
   \caption{Autocorrelation of the slightly smoothed power spectrum. The large spacing, as well as separations between $\ell$\,=\,0 and $\ell$\,=\,1 modes are clearly present.}
              \label{auto}%
    \end{figure}

The first step is to measure the large spacing that should appear at least between radial modes.
The power spectrum is autocorrelated to search for periodicity.
Each peak of this autocorrelation (see Fig.~\ref{auto}) corresponds to a structure present in the power spectrum.    
One of the three strong groups of peaks at about 1.7, 3.5, and 5.2\,$\mu$Hz should correspond to the large spacing.
By visually inspecting the power spectrum, the value
of about 3.5\,$\mu$Hz is adopted as the large separation, the two others are spacings between the $\ell$\,=\,0 and $\ell$\,=\,1 modes. This large separation value is
in good agreement with the scaled value from the solar case (see Sect.~ \ref{lse}).

\subsection{Extraction of mode parameters}
   \begin{figure}
   \resizebox{\hsize}{!}{\includegraphics{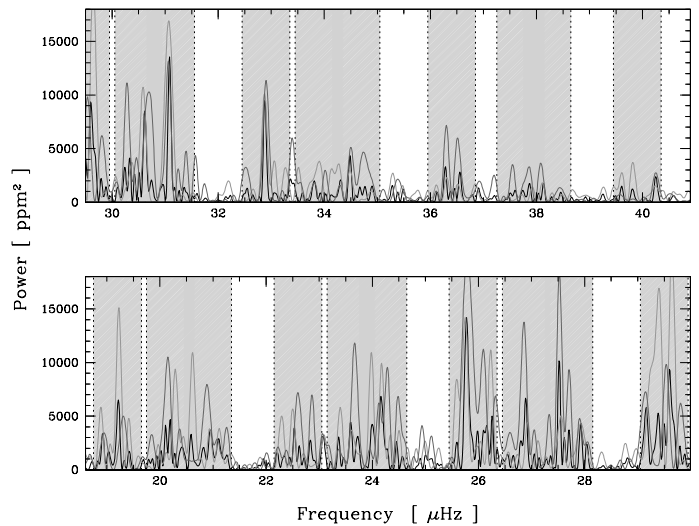}}
   \caption{Power spectra of the whole dataset (black) and of the two half-long subsets (grey). The frequencies of the oscillation peaks
   change from one subset to the other, which is a sign of a finite lifetime. The initial guess for the identification of modes are indicated
   by shaded regions: these regions are regularly spaced in agreement with the large separation deduced from the autocorrelation. 
   Every second region has a structure more simple and narrow, 
   corresponding to our identification of $\ell$\,=\,1 modes,
   the others are more complex and wide and correspond to $\ell$\,=\,0 and~2 modes.}
              \label{fit2}%
    \end{figure}
   \begin{figure*}
   \resizebox{\hsize}{!}{\includegraphics{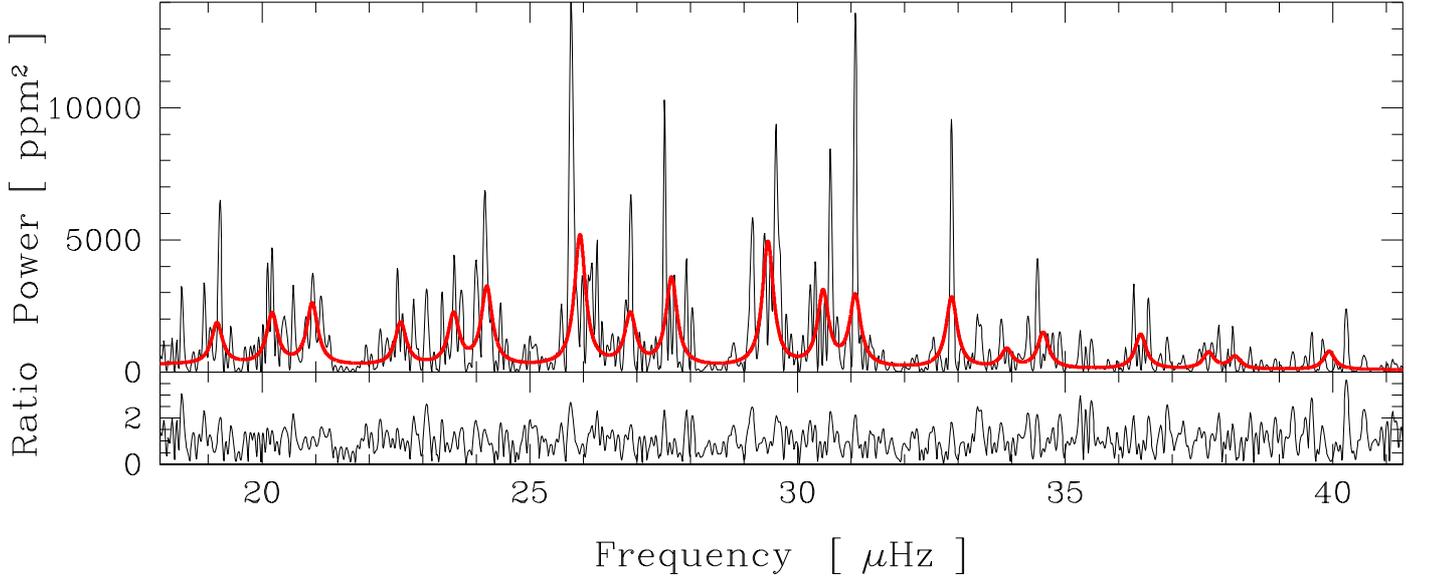}}
   \caption{{\bf Top:} Lorentzian fit (thick red line) of the observed power spectrum assuming the same lifetime for all the oscillation modes.
   {\bf Bottom:} Ratio of the observed spectrum over the Lorentzian fit (in amplitude). As expected, this ratio does not show any correlation with the fit.}
              \label{fit}%
    \end{figure*}
The power spectra in Figs.~\ref{fit2} and~ \ref{fit} clearly exhibit a regularity that allows us to identify $\ell$\,=\,0 to $\ell$\,=\,2 modes. 
Each mode consists of several peaks, which is the clear signature of a finite lifetime shorter than the observing time span. In order to 
determine the mode frequencies, as well as amplitude and lifetime of the modes, we fitted the power
spectrum using a maximum likelihood estimation (MLE) method.
MLE has been applied widely in the helioseismic community (e.g., see Schou \cite{schou}, Appourchaux et
al. \cite{appour1}, and Chaplin et al. \cite{chaplin}).
Our program uses the IDL routines developed by T. Appourchaux (\cite{appour1}). This method has already been used with success for the red giant $\epsilon$~Oph (Barban et al. \cite{barban2}).

The modelled power spectrum for a series of M oscillation modes, P($\nu_k$), is
\begin{equation}
P ( \nu_k ) = \sum_{n=1}^M \left( H_n \frac{1}{1+\left(\frac{2 ( \nu_k - \nu_n)^2}{\Gamma_n}\right)} \right) + B,
\end{equation}
where $H_n$ is the height of the Lorentzian profile, $\nu_n$ is the oscillation mode frequency, $\Gamma_n$
is the mode line-width, and $B$ is the background noise.
The fit was performed on the non-over-sampled power spectrum to minimize the interdependency of the points. The quantity
that was minimized is
\begin{equation}
L = \sum_{k=1}^K \left( \ln P(\nu_k) + \frac{P_{obs}(\nu_k)}{P(\nu_k)} \right) ,
\end{equation}
where $K$ is the number of bins, i.e., the number of Fourier frequencies.

\begin{figure}
   \resizebox{\hsize}{!}{\includegraphics{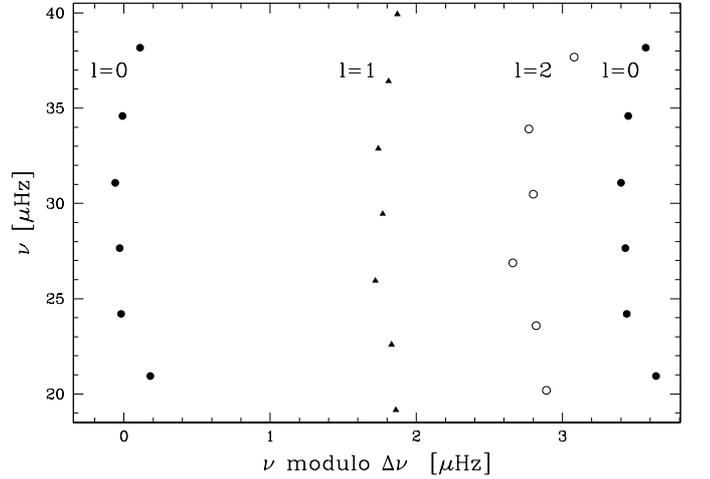}}
   \caption{Echelle diagram of identified modes with a large separation of $\Delta\nu_0$\,=\,3.45\,$\mu$Hz. The modes $\ell$\,=\,0 ($\bullet$),
     $\ell$\,=\,1 ($\blacktriangle$), and $\ell$\,=\,2 ($\circ$) follow ridges, but $\ell$\,=\,1 modes are
     situated too far to the right to
    follow the asymptotic relation.}
              \label{dech}%
    \end{figure}

Only a few peaks belong to a given mode, indicating a lifetime longer than 10 days.
It is thus also difficult to derive the correct Lorentzian shape for each mode.
Therefore some parameters were fixed to avoid incorrect parameter determinations:
\begin{itemize}
\item The noise was determined independently of the MLE method (see Sect.~\ref{nb}).
\item Since no difference between the shapes of modes of different degree $\ell$ is detected and the value
of v\,$\sin i$ is extremely small, resulting in an 
expected rotational splitting value smaller than 0.03\,$\mu$Hz, we assumed that it is zero.
\item When fitting all modes without fixing the width of the Lorentzian envelope, we clearly saw that the fit was not robust enough to 
provide an accurate determination of all mode parameters.
The method was thus first to find a mean value for the Lorentzian width and to fix this mean value for all modes.
The mean width of the Lorentzian is obtained by individually fitting all modes and by taking their mean,
rejecting the values that obviously correspond to a poor fit.
The determined mean
is 0.25\,$\pm$\,0.06\,$\mu$Hz, which 
corresponds to a mode lifetime of 14.7$^{+4.7}_{-2.9}$\,d.
The use of sub-series show the stochastic nature of the mode excitation caused by the different fine structure of peaks in each spectrum, and give an indication of their width.
We note that we also checked that no significant width difference was found for modes with different
degrees $\ell$ (by comparing their mode-width means).
\end{itemize}
All modes  were then fitted with fixed parameters as deduced above. We note that the initial guess values are indicated by shaded regions in Fig.~\ref{fit2}.
These regions are already in good agreement with a regular spacing between modes.
Every second region has a structure that is more simple and narrow, corresponding to our identification of $\ell$\,=\,1 modes,
the others are more complex and wide. In this last case, we needed to fit two modes per region to reproduce the power spectrum (which correspond to $\ell$\,=\,0 and~2 modes).
At 38\,$\mu$Hz, the signal-to-noise ratio is far smaller and it becomes more difficult to differentiate
between modes: for the sake of homogeneity, we also decided to fit two modes in this region.
The frequencies of these last two modes will however be more uncertain.
The result is listed in Table~\ref{tab1}. 
The formal uncertainties associated with MLE are well understood, as explained
by Libbrecht (\cite{lib}) and Toutain \& Appourchaux (\cite{toutain}).
The echelle diagram with the nineteen identified modes is shown in Fig.~\ref{dech}. At higher and lower frequency (above 40\,$\mu$Hz and below 19\,$\mu$Hz), the amplitude of the modes is either too small or the noise too high to unambiguously identify additional modes.
The ratio of the observed power spectrum to the fit is shown in Fig.~\ref{fit}: it appears to be pure
noise and has a mean value of 1. Moreover, no correlation was found between the amplitude of this ratio and the amplitude of the fit.
We note that the non-radial modes are as well aligned as radial or non-mixed modes.
However, the separation between $\ell$\,=1 and $\ell$\,=0 modes is not fully compatible with
the asymptotic relation. The $\ell$\,=1 modes are indeed too far to the right in the echelle diagram. 

\begin{figure}
   \resizebox{\hsize}{!}{\includegraphics{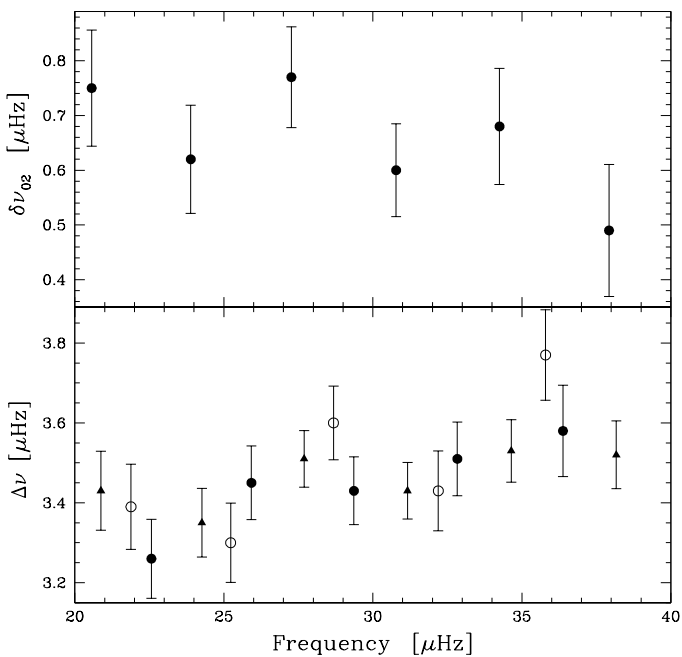}}
   \caption{{\bf Top:} Small spacing versus frequency. 
    {\bf Bottom:} Large spacing versus frequency. The variations of the large separation with frequency show a clear oscillation.
    The symbols used are the same as for Fig.~\ref{dech}.}
              \label{diagastero}%
    \end{figure}

The large and small separations are shown in Fig.~\ref{diagastero}. The mean large separation has a value of $\Delta\nu$\,=\,3.47\,$\pm$\,0.12\,$\mu$Hz, and the values for different
degrees $\ell$, $\Delta\nu_{\ell}$, are:
$\Delta\nu_0$\,=\,3.45\,$\pm$\,0.12\,$\mu$Hz, $\Delta\nu_1$\,=\,3.46\,$\pm$\,0.07\,$\mu$Hz, and $\Delta\nu_2$\,=\,3.50\,$\pm$\,0.19\,$\mu$Hz. 
We can identify a small oscillation of the large spacing that varies with frequency, which is a clear signature of the
second helium ionization zone (see e.g., Monteiro \& Thompson \cite{monteiro}).
The small separation has a mean value
of $\delta\nu_{02}$\,=\,0.65\,$\pm$\,0.10\,$\mu$Hz and seems to decrease with frequency. 
The small value of the frequency difference between these modes make their frequency determination more uncertain than that of 
$\ell$\,=\,1 modes, which are not affected by neighbouring modes.

\begin{table}
\caption[]{Frequency and amplitude of identified modes.
The uncertainties in the last given digit of the frequencies are noted in parentheses. For the S/N estimates
(see Barban et al. \cite{barban2}), the signal is
taken to be the height of the fitted Lorentzian profile and the noise is determined according to the procedure described in Sect.~\ref{nb}.}
\begin{center}
\begin{tabular}{ccccc}
\hline
\hline
Degree  & Frequency & Mode height & S/N & Amplitude \\
$\ell$ & $\mu$Hz &ppm$^2$ * 1000 / $\mu$Hz& & ppm \\
\hline
0  & 20.94 (7)& 7.8&  10.8&79\,$\pm$\,15\\
0  & 24.20 (7)& 10.0& 17.6&89\,$\pm$\,15\\
0 & 27.65  (6)& 11.3& 24.8&94\,$\pm$\,16\\
0 & 31.08  (6)& 9.0&  23.7&84\,$\pm$\,14\\
0 & 34.59  (7)& 4.6&  14.0&60\,$\pm$\,11\\
0 & 38.17  (9)& 1.6 & 5.6&35 \,$\pm$\,8\\
1  & 19.16 (7)& 5.3&  6.2&65\,$\pm$\,14\\
1  & 22.59 (7)& 5.5&  8.6&66\,$\pm$\,13\\
1  & 25.94 (5)& 16.8& 33.1&115\,$\pm$\,19\\
1 & 29.45  (5)& 16.0& 38.8&112\,$\pm$\,18\\
1 & 32.88  (5)& 9.1&  26.0&85\,$\pm$\,14\\
1 & 36.41  (6)& 4.4&  14.6&59\,$\pm$\,11\\
1 & 39.93  (6)& 2.3 & 8.8&43 \,$\pm$\,9\\
2  & 20.19 (8)& 6.5&  8.4&71\,$\pm$\,14\\
2  & 23.58 (7)& 6.5&  10.9&71\,$\pm$\,13\\
2  & 26.88 (7)& 6.5&  13.7&72\,$\pm$\,13\\
2 & 30.48  (6)& 9.5&  24.1&86\,$\pm$\,15\\
2 & 33.91  (8)&  2.3& 7.0&43 \,$\pm$\,9\\
2 & 37.68  (8)&  2.1& 7.3&41 \,$\pm$\,8\\
\hline
\hline
\end{tabular}\\
\end{center}
\label{tab1}
\end{table}
 \subsection{Oscillation amplitudes}
\begin{figure}
   \resizebox{\hsize}{!}{\includegraphics{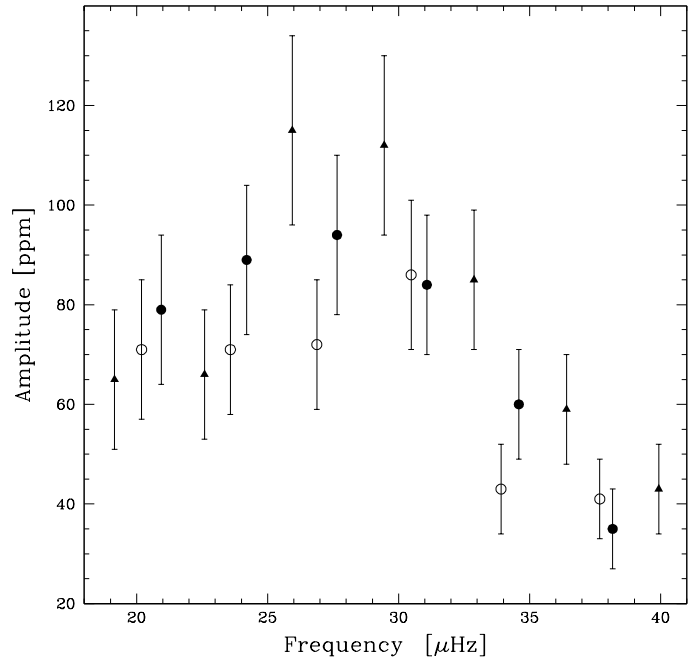}}
   \caption{Amplitude of the oscillation modes versus frequency. The symbols used are the same as for Fig.~\ref{dech}.}
              \label{ampli}%
    \end{figure}

The fit of the Lorentzian profiles to the power spectrum infers the height of all oscillation modes. 
Since the modes are resolved and because of the normalization of the power spectrum, the {\sc rms}
amplitude is measured to be (see Baudin et al. \cite{baudin})
\begin{equation}
A = \sqrt{H \pi \Gamma} ,
\end{equation}
where H and $\Gamma$ are the height and width (FWHM) of the Lorentzian function, respectively, in the power density spectrum. The amplitudes are in the range of 35 -- 115\,ppm (see 
Fig.~\ref{ampli}). The error bars are derived from the Hessian matrix  
and the correlation between the height and width has been taken into account (Toutain \& Appourchaux \cite{toutain}, Appourchaux, private communication).
We note that the amplitudes of the $\ell$\,=\,1 modes are the highest. 

As noticed by Kjeldsen et al. (\cite{kjel}), measurements made on different stars with different instruments using different techniques, in different
spectral lines or bandpasses, have different sensitivity to the oscillations. It is thus important to
derive a bolometric amplitude that is independent of the
instrument used. We computed the maximum bolometric amplitude of the $\ell$\,=\,0 modes, because their
visibility coefficients do not depend on the inclination of the star.
According to Michel et al. (\cite{michel}), who derived the CoRoT response for radial modes, radial-mode
amplitudes of HR~7349 must be divided by 1.16 to obtain 
the bolometric amplitudes. We find  that A$_{\rm bol, \ell=0, max}$\,=\,81\,ppm, which corresponds to 32 times the solar value (Michel et al. \cite{michel}). 
The scaling laws for both the large separation and the frequency of the maximum amplitude (Kjeldsen \& Bedding, \cite{kjebed}), coupled with the non-asteroseismic constraints,
infer
a mass for HR~7349 of about 1.2\,M$_{\odot}$
The derived amplitude is in good agreement with a scaling function $( L / M )^{s} / \sqrt{T_{\rm eff}/T_{\rm eff,\odot}}$, with $s$ close to 0.8, 
which is in-between the values given by Samadi et al. (\cite{samadi}; s=0.7) and Kjeldsen \& Bedding (\cite{kjebed}; s=1).

\section{Conclusion}
The red giant star HR~7349 has been observed for about 156 days by the CoRoT satellite.
These observations have yielded a clear detection of p-mode oscillations. 
As already mentioned by De Ridder et al. (\cite{joris2}), non-radial modes are observed in red giants.
Nineteen identifiable modes of degree $\ell$\,=\,0 to $\ell$\,=\,2 appear
in the power spectrum between 18 and 42\,$\mu$Hz with average large and small spacings of
3.47 and 0.65\,$\mu$Hz, respectively, and a maximum bolometric amplitude of 81\,ppm. We note that the amplitude of $\ell$\,=\,1 modes is even larger
than that of the radial modes.

All modes of the same degree are aligned in the echelle diagram, which is a sign of modes that follow the asymptotic relation. 
However, the separation between $\ell$\,=1 and $\ell$\,=0 modes is not fully compatible with
this asymptotic relation. All frequency patterns
are theoretically expected for red giants (Dupret et al. \cite{dupret}), from very complex to regular:
our observations correspond to a red giant for which the radiative
damping of non-radial modes is large and only radial and non-radial modes completely trapped in the envelope can be observed.

By fitting Lorentzian profiles to the power spectrum, it has also been possible to unambiguously derive, for the first time for a red giant, 
a mean line-width of 0.25\,$\mu$Hz corresponding to a mode lifetime of 14.7\,days. This lifetime
is in agreement with the scaling law $T_{\rm eff}^{-4}$ suggested by Chaplin et al.~\cite{chaplin9},
although is a little too long. This relation however has 
yet to be verified for a larger number of red giants with different physical properties.
The theoretical study of this red giant, including asteroseismic and non-asteroseismic constraints, will be the subject of a second paper.

\begin{acknowledgements}
FC is a postdoctoral fellow of the Fund for Scientific Research, Flanders (FWO). AM is a postdoctoral researcher of the 'Fonds de la recherche scientifique' FNRS, Belgium.
TK is supported by the Canadian Space Agency
and the Austrian Science Found (FWF). The research leading to these results has received funding from the
Research Council of K.U.Leuven under grant agreement GOA/2008/04, from
the Belgian PRODEX Office under contract C90309: CoRoT Data Exploitation,
and from the FWO-Vlaanderen under grant O6260. We thank T. Appourchaux for helpful comments.
\end{acknowledgements}



\begin{thebibliography}{}
\bibitem[2004]{aigrain} Aigrain, S., Favata, F., \& Gilmore, G. 2004, A\&A, 414, 1139
\bibitem[1999]{alonso} Alonso, A., Arribas, S., Martinez-Roger, C. 1999, A\&AS, 140, 261
\bibitem[1998]{andersen} Andersen, B., Appourchaux, T., \& Crommelnynck, D. 1998, in Sounding solar and stellar interiors, ed. Provost \& Schmider, 181, 147
\bibitem[2008]{appourchaux} Appourchaux, T., Michel, E., Auvergne, M., et al. 2008, A\&A, 488, 705
\bibitem[1998]{appour1} Appourchaux, T., Gizon, L., \& Rabello-Soares, M.C. 1998, A\&AS, 132, 107
\bibitem[1992]{arenou} Arenou, F., Grenon, M., \& Gomez, A. 1992, A\&A, 258, 104
\bibitem[2009]{auvergne} Auvergne, M., Bodin, P., Boisnard, L., et al. 2009, A\&A, accepted
\bibitem[2006]{bag} Baglin, A. 2006, ESA SP-1306, The CoRoT mission, Fridlund, Baglin, Lochard and Conroy eds, Noordwijk, The Nederlands
\bibitem[2007]{barban2} Barban, C., Matthews, J.M., De Ridder, J., et al. 2007, A\&A, 468, 1033
\bibitem[2004]{barban} Barban, C., De Ridder, J., Mazumdar, A., et al. 2004, ESA SP-559, SOHO 14
Helio- and Asteroseismology, ed. D. Danesy, 113
\bibitem[2005]{baudin} Baudin, F., Samadi, R., Goupil, M.-J., et al. 2005, A\&A, 433, 349
\bibitem[2007]{bk07} Bedding, T.R., \& Kjeldsen H. 2007, CoAst, 150, 106
\bibitem[2008]{bu08} Burki, G., et al. 2008, $\mathtt{http://obswww.unige.ch/gcpd/ph13.html}$ 
\bibitem[2008]{cel08} Carrier, F., Eggenberger, P., \& Leyder, J.C. 2008, JPHCS, 118, 2047
\bibitem[2009]{chaplin9} Chaplin, W.J., Houdek, G., Karoff, C., et al. 2009, A\&A, 500, L21
\bibitem[2006]{chaplin} Chaplin, W.J., Appourchaux, T., Baudin, F., et al. 2006, MNRAS, 369, 985
\bibitem[2004]{chris} Christensen-Dalsgaard, J. 2004, Sol. Phys., 220, 137
\bibitem[1995]{cd95} Christensen-Dalsgaard, J., Bedding, T.R., \& Kjeldsen, H.\ 1995, ApJ, 443, L29
\bibitem[2009]{dupret} Dupret, M.-A., Belkacem, K., Samadi, R., et al. 2009, A\&A in press
\bibitem[2009]{joris2} De Ridder, J., Barban, C., Baudin., F., et al. 2009, Nature, 459, 398
\bibitem[2006]{joris1} De Ridder, J., Barban, C., Carrier, F., et al. 2006, A\&A, 448, 689
\bibitem[2003]{fm03} Fernandes, J., \& Monteiro, M.J.P.F.G. 2003, A\&A, 399, 243
\bibitem[1996]{flower} Flower, P. 1996, ApJ, 469, 355
\bibitem[2002]{frandsen} Frandsen, S., Carrier, F., Aerts, C., et al. 2002, A\&A, 394, L5
\bibitem[2005]{kjel} Kjeldsen, H., Bedding, T., Butler, R.P., et al. 2005, ApJ, 635, 1281
\bibitem[1995]{kjebed} Kjeldsen, H., \& Bedding, T. 1995, A\&A, 293, 87
\bibitem[1985]{harvey} Harvey, J. 1985, in Future Missions in Solar, Heliospheric, \& Space Plasma Physics, ed. Rolfe \& Battrick, ESA SP, 235, 199
\bibitem[2009]{hekker} Hekker, S., Kallinger, T., Baudin, F., et al. 2009, A\&A, accepted
\bibitem[2007]{hekker1} Hekker, S., \& Mel\'endez, J. 2007, \aap, 475, 1003
\bibitem[1998]{le98} Lejeune, T., Cuisinier, F., \& Buser, R. 1998, A\&AS, 130, 65
\bibitem[1992]{lib} Libbrecht, K.G. 1992, ApJ, 387, 712
\bibitem[2009]{michel} Michel, E., Samadi, R.,Baudain, F., et al. 2009, A\&A 495, 979 
\bibitem[1998]{monteiro} Monteiro, M.J.P.F.G., \& Thompson, M.J. 1998, IAU Symposium, 185, 317
\bibitem[2007]{samadi} Samadi, R., Georgobiani, D., Trampedach, R., et al. 2007, A\&A, 463, 297
\bibitem[2002]{santos} Santos, N.C., Mayor, M., Naef, D., et al. 2002, A\&A, 392, 215
\bibitem[1992]{schou} Schou, J. 1992, PhD thesis, Aarhus Universitet, Denmark
\bibitem[2006]{stello} Stello, D., Kjeldsen, H., Bedding, T.R., Buzasi, D. 2006, A\&A, 448, 709
\bibitem[2008]{tarrant} Tarrant, N.J., Chaplin, W.J., Elsworth, Y.P., et al. 2008, CoAst, 157, 92
\bibitem[1980]{tassoul80} Tassoul, M. 1980, ApJS, 43, 469
\bibitem[1994]{toutain} Toutain, T., \& Appourchaux, T. 1994, A\&A, 289, 649
\bibitem[2007]{vanleeuwen} van Leeuwen, F. 2007, A\&A, 474, 653
\end{thebibliography}
\end{document}